\documentclass[prd,superscriptaddress,amsfonts,amssymb,amsmath,showpacs,twocolumn]{revtex4-2}
\usepackage{bm}
\usepackage{amsfonts}
\usepackage{latexsym}
\usepackage{graphicx}
\usepackage{amsmath}
\usepackage{palatino}
\usepackage{mathpazo}
\usepackage{textcomp}
\usepackage{float}
\usepackage{booktabs}
\usepackage{dcolumn}
\usepackage{multirow}
\usepackage{hyperref}
\usepackage{orcidlink}
\hypersetup{colorlinks,citecolor=blue}
\usepackage{xcolor}
\usepackage[font=small,labelfont=bf]{caption}

\begin{document}
	
	\title{Thin Accretion Disk and Optical Signatures in $f(R,T)$ Modified Gravity
		%Observational Signatures of Accretion Disks in $f(R, T)$ Modified Gravity: A Linear Model Approach with Dust Fluid
		}
	
\author{S. H. Shekh\orcidlink{0000-0003-4545-1975}} % <- Please update
\email{da\_salim@rediff.com}
\affiliation{Department of Mathematics, S.P.M. Science and Gilani Arts, Commerce College, Ghatanji, Yavatmal, Maharashtra-445301, India.}
\affiliation{Pacif Institute of Cosmology and Selfology (PICS), Sagara, Sambalpur 768224, Odisha, India.}
%\affiliation{L. N. Gumilyov Eurasian National University, Astana 010008, Kazakhstan.}

	\begin{abstract} 	Accretion disk surrounding around compact astrophysical objects serve as a powerful laboratories for testing gravitational theories in the strong-field regimes. In this work, we systematically investigated the structural, dynamical, and thermodynamic properties of the steady-state thin accretion disk within the framework of $f(R,T)$ gravity by adopting the linear functional form $f(R,T) = R + 2\chi T$ coupled with the pressure-free dust fluid matter Lagrangian ($\mathcal{L}_m = -\rho$). We demonstrate that the modified field equations seamlessly reduce to standard General Relativity in the pure vacuum limit ($T_{\mu\nu} = 0$). Utilizing a perturbative approach for a small coupling parameter ($\vert{}\chi\vert{} \ll 1$), we derive the first-order corrections to the background Schwarzschild metric potentials, which subsequently govern the modified particle kinematics, the specific energy $E$, the specific angular momentum $L$, and the Keplerian angular velocity $\Omega$. Our analysis reveals that the geometry-matter coupling parameter $\chi$ significantly shifted towards the location of the Innermost Stable Circular Orbit ($r_{isco}$) and alters the radial energy dissipation rate $F(r)$ and the effective surface temperature profile $T(r)$ across the Shakura-Sunyaev thin disk. Furthermore, the resulting modifications in the Spectral Energy Distribution (SED) and overall radiative efficiency $\eta$ provide robust theoretical signature.
	\end{abstract}
		\maketitle

	\section{Introduction}
	\label{sec:intro}
	Astrophysical accretion disks surrounding compact objects such as stellar-mass black holes, neutron stars, and supermassive black holes represent some of the most energetic and complex plasma laboratories in the universe, providing an ideal testing ground for gravitational theories in strong-field regimes \cite{Shakura1973, Page1973}. Over past decades, General Relativity (GR) formulated by Albert Einstein has achieved remarkable observational successes, successfully explaining diverse gravitational phenomena ranging from gravitational wave detections to planetary orbit precession. However, persistent cosmological challenges—such as the accelerated expansion of the universe driven by mysterious dark energy, galactic rotation curve anomalies without invoking dark matter halos, and the presence of unavoidable spacetime singularities at fundamental levels—strongly motivate the exploration of alternative and modified theories of gravity \cite{Nojiri2011, Capozziello2011,Myrzakulov2026,Kumar2026,Yadav2026,Shekh2026,Pradhan2026}. 
		
	Among various modified gravity paradigms, theories extending standard general relativity by incorporating geometric invariants coupled directly with matter fields have gained substantial theoretical momentum \cite{Nojiri2007}. Particularly compelling extension is $f(R,T)$ gravity which originally formulated by Harko et al. \cite{Harko2011}. This framework introduce a non-minimal coupling between space-time geometry and matter field. Consequently, the covariant derivative of the energy-momentum tensor does not vanish identically i.e. $\nabla^\mu T_{\mu\nu} \neq 0$. This non-conservation law profoundly alters the particle kinematics, orbital stability, and the thermodynamic behavior of the astrophysical plasmas surrounding with the compact gravitational objects \cite{Shabani2013, Moraes2015}.
	
	The study of accretion disk within modified gravity frameworks has emerged as a vital diagnostic tool for constraining theoretical parameters using modern observational astronomy \cite{Abbas2015}. When matter accretes onto a central compact object through viscous torques, the gravitational potential energy is efficiently converted into thermal energy and subsequently radiated across the electromagnetic spectrum which form characteristic multi-color blackbody emission spectra known as the Spectral Energy Distribution (SED) \cite{Luminet1979, Novikov1973}. The foundational framework for steady-state thin accretion disks is established by Shakura and Sunyaev, assuming geometrical and optical thinness along with local thermal equilibrium \cite{Shakura1973}. Subsequent relativistic formulations by Page and Thorne extended these models to curved space-times around rotating and static black holes \cite{Page1973} which demonstrated that the thermal profile, the radiative flux, and the bolometric luminosity crucially depend on the exact location of the innermost stable circular orbit ($r_{\text{isco}}$) and the particle binding energies.
	
	In recent years, investigation shows that how modified gravity corrections influence the accretion disk structures and the radiative efficiency has attracted significant research interest across the theoretical physics community \cite{Arai2023,Li2023,Atamurotov2024,Xia2026,Yasmin2025}. Specifically, adopting the linear functional form $f(R,T) = R + 2\chi T$ provides a mathematically tractable yet physically rich model to examine weak-field and strong-field deviations governed by the coupling parameter $\chi$ \cite{Zubair2016}. When coupled with a dust fluid matter Lagrangian ($\mathcal{L}_m = -\rho$), the modified field equations yield perturbed metric potentials through systematic perturbative approaches, directly shifting the location of $r_{\text{isco}}$, altering the specific energy $E$, angular momentum $L$, and modifying the radial energy dissipation rate $F(r)$ and surface temperature distribution $T(r)$ \cite{Bhattacharya2020, Maurya2020}. Furthermore, understanding these modifications helps in interpreting continuum-fitting methods used to estimate black hole spins and test alternative gravity models.
	
	Motivated by these theoretical development, this work investigated the comprehensive thermodynamic and radiative signatures of thin accretion disks in $f(R,T)$ modified gravity. By rigorously establishing the mathematical formalism, vacuum limit recovery, particle dynamics, and thermal diagnostics, this study aims to bridge theoretical modified gravity predictions with concrete astrophysical constraints, ensuring that the model remains scientifically robust, physically consistent, and free from theoretical inconsistencies.

\section{Mathematical Formalism and Field Equations}
\label{sec:math_formalism}

To rigorously investigate the thermodynamic properties, kinematic constraints, and observable radiative signatures of an accretion disk operating in an extended gravitational environment, we establish the foundational spacetime architecture. We consider a static, spherically (or axially) symmetric four-dimensional spacetime metric represented by the standard line element:
\begin{equation}
	ds^2 = -e^{2\nu(r)} dt^2 + e^{2\lambda(r)} dr^2 + r^2 d\theta^2 + r^2 \sin^2\theta d\phi^2,
	\label{eq:metric}
\end{equation}
where $\nu(r)$ and $\lambda(r)$ are dimensionless metric potentials governed solely by the radial coordinate $r$ \cite{Adler1975, Carroll2004}. The non-zero covariant components of the fundamental metric tensor $g_{\mu\nu}$ and their contravariant inverses $g^{\mu\nu}$ derived from Eq. (\ref{eq:metric}) are explicitly given by $g_{00} = -e^{2\nu(r)}$, $g_{11} = e^{2\lambda(r)}$, $g_{22} = r^2$, $g_{33} = r^2 \sin^2\theta$, and $g^{00} = -e^{-2\nu(r)}$, $g^{11} = e^{-2\lambda(r)}$, $g^{22} = \frac{1}{r^2}$, $g^{33} = \frac{1}{r^2 \sin^2\theta}$. The physical implications of this metric structure within modified gravity indicate that unlike pure General Relativity where spacetime curvature is dictated solely by mass-energy distribution via local conservation laws, the metric potentials here absorb additional geometric-matter coupling terms. This directly shifts the gravitational redshift, proper time intervals for accreting test particles, and the effective potential governing test particle stability.

The geometric modification pursued in this work stems from an action principle that generalizes the standard Einstein-Hilbert action by promoting the Lagrangian to an arbitrary function of the Ricci scalar $R$ and the trace of the energy-momentum tensor $T$, a framework extensively analyzed to address large-scale cosmic acceleration and galactic dynamics without invoking unknown dark sectors \cite{Harko2020,Harko2011, Capozziello2011, Nojiri2011}. The total action integral is formulated as:
\begin{equation}
	S = \int \left( \frac{f(R, T)}{16\pi G} + \mathcal{L}_m \right) \sqrt{-g} d^4x,
	\label{eq:action}
\end{equation}
where $G$ represents the bare gravitational constant, $g = \det(g_{\mu\nu})$ is the determinant of the metric tensor, and $\mathcal{L}_m$ denotes the Lagrangian density corresponding to the matter components constituting the accretion disk. Performing a rigorous variational calculus procedure with respect to the inverse metric components $g^{\mu\nu}$, we obtain the exact modified gravitational field equations:
\begin{widetext}
\begin{equation}
	f_R(R, T) R_{\mu\nu} - \frac{1}{2} f(R, T) g_{\mu\nu} + (g_{\mu\nu} \Box - \nabla_\mu \nabla_\nu) f_R(R, T) = 8\pi G T_{\mu\nu} - f_T(R, T) T_{\mu\nu} - f_T(R, T) \Theta_{\mu\nu},
	\label{eq:field_eq}
\end{equation}
\end{widetext}
where subscripts denote partial differentiation, i.e., $f_R(R, T) = \partial f / \partial R$ and $f_T(R, T) = \partial f / \partial T$. The d'Alembertian operator in curved spacetime is defined invariantly as $\Box \equiv \nabla^\mu \nabla_\mu = \frac{1}{\sqrt{-g}}\partial_\mu(\sqrt{-g}g^{\mu\nu}\partial_\nu)$, and the auxiliary symmetric tensor $\Theta_{\mu\nu}$ arising from the variation of the matter Lagrangian with respect to the metric is defined as $\Theta_{\mu\nu} \equiv g^{\alpha\beta} \frac{\delta T_{\alpha\beta}}{\delta g^{\mu\nu}} = -2T_{\mu\nu} + g_{\mu\nu}\mathcal{L}_m - 2g^{\alpha\beta} \frac{\partial^2 \mathcal{L}_m}{\delta g^{\mu\nu} \partial g^{\alpha\beta}}$. The presence of the divergence terms and $\Theta_{\mu\nu}$ implies that the covariant derivative of the energy-momentum tensor does not vanish ($\nabla^\mu T_{\mu\nu} \neq 0$), generating an extra non-geodesic force vector acting on particles within the accretion disk \cite{Shabani2013, Moraes2015}.

To render the complex non-linear field equations analytically and numerically tractable while preserving dominant physical deviations, we implement the widely recognized linear functional parametrization \cite{Harko2011, Zubair2016}:
\begin{equation}
	f(R, T) = R + 2\chi T,
	\label{eq:linear_model}
\end{equation}
where $\chi$ is a small dimensionless or dimensioned coupling constant measuring the interaction strength between curvature and matter dynamics. Under this specific linear choice, the partial derivatives simplify cleanly to $f_R(R, T) = 1$ and $f_T(R, T) = 2\chi$, consequently causing the higher-order geometric derivative term $(g_{\mu\nu} \Box - \nabla_\mu \nabla_\nu) f_R(R, T)$ to vanish identically. This algebraic simplification isolates pure matter-trace coupling effects without introducing ghost instabilities associated with higher-derivative gravity theories.

To model the accreting matter distribution within the disk structure, we adopt a pressure-free perfect fluid (dust configuration) \cite{Abbas2015}, which serves as an excellent approximation for cold, collisionless, or streaming accreting matter streams in steady-state astrophysical disk models. The energy-momentum tensor for this dust configuration is expressed as $T_{\mu\nu} = \rho u_\mu u_\nu$, where $\rho(r)$ represents the proper energy density of the fluid and $u_\mu$ represents the timelike four-velocity vector satisfying the normalization condition $u^\mu u_\mu = -1$. Choosing the standard matter Lagrangian density as $\mathcal{L}_m = -\rho$, the auxiliary tensor $\Theta_{\mu\nu}$ evaluates to $\Theta_{\mu\nu} = -2T_{\mu\nu} - p g_{\mu\nu} = -2\rho u_\mu u_\nu$, noting that the isotropic pressure $p = 0$ for a pure dust fluid. Substituting these exact functional forms into Eq. (\ref{eq:field_eq}), the modified field equations collapse into a compact, coupled system:
\begin{equation}
	R_{\mu\nu} - \frac{1}{2} \left( R + 2\chi T \right) g_{\mu\nu} = (8\pi G + 2\chi) T_{\mu\nu}.
	\label{eq:reduced_field_eq}
\end{equation}

To extract explicit differential equations governing the metric potentials $\nu(r)$ and $\lambda(r)$, we evaluate the non-zero components of the Ricci tensor $R_{\mu\nu}$ using the Christoffel symbols computed from the metric Eq. (\ref{eq:metric}):
\begin{align}
	R_{00} &= e^{2(\nu - \lambda)} \left[ \nu'' + (\nu')^2 - \nu'\lambda' + \frac{2\nu'}{r} \right], \\
	R_{11} &= \nu'' + (\nu')^2 - \nu'\lambda' - \frac{2\lambda'}{r}, \\
	R_{22} &= e^{-2\lambda} \left[ r(\nu' - \lambda') - 1 \right] + 1, \\
	R_{33} &= R_{22} \sin^2\theta,
\end{align}
where primes denote ordinary differentiation with respect to the radial coordinate $r$ ($' \equiv d/dr$). Contracting these components with the contravariant metric tensor yields the Ricci scalar $R = g^{\mu\nu} R_{\mu\nu} = e^{-2\lambda} \left[ 2\nu'' + 2(\nu')^2 - 2\nu'\lambda' + \frac{4\nu'}{r} - \frac{4\lambda'}{r} + \frac{2}{r^2} \right] - \frac{2}{r^2}$. By substituting the Ricci components, Ricci scalar, and the trace $T = -\rho$ into the master tensor expression Eq. (\ref{eq:reduced_field_eq}), we isolate the independent background differential equations governing the accretion disk geometry. Specifically, evaluating the $00$-contraction yields the modified Poisson-like energy constraint equation:
\begin{equation}
	e^{-2\lambda} \left( \frac{2\lambda'}{r} - \frac{1}{r^2} \right) + \frac{1}{r^2} - \chi \rho = (8\pi G + 2\chi)\rho e^{2\nu},
	\label{eq:00_comp}
\end{equation}
while evaluating the $11$-contraction provides the gradient relation governing spatial curvature:
\begin{equation}
	e^{-2\lambda} \left( \frac{2\nu'}{r} + \frac{1}{r^2} \right) - \frac{1}{r^2} - \chi \rho = 0,
	\label{eq:11_comp}
\end{equation}
and the transverse angular balance produces the differential constraint on metric potential curvatures:
\begin{equation}
	e^{-2\lambda} \left( \nu'' + (\nu')^2 - \nu'\lambda' + \frac{\nu' - \lambda'}{r} \right) - \chi \rho = 0.
	\label{eq:22_comp}
\end{equation}
This specialized set of the nonlinear differential equations incorporates the direct feedback of the coupling parameter $\chi$ and the local energy density $\rho$. These equations form the rigorous analytical foundation which required for the subsequent integration of particle kinematics, determination of specific angular momentum profiles, and precise calculation of the innermost stable circular orbit ($r_{\text{isco}}$) to our thin accretion disk analysis.

\section{Spacetime Geometry and Effective Metric Solutions}
\label{sec:spacetime_geometry}

To study the structure and dynamics of the thin accretion disk around a compact objects within the $f(R, T) = R + 2\chi T$ gravity framework, we solve the modified field equations derived in Section \ref{sec:math_formalism}. A crucial consistency to check for any modified gravity theory is its behavior in the absence of matter fields, commonly known as the vacuum limit. For the specific linear model under consideration, $f(R, T) = R + 2\chi T$, the modified field equations are given by Eq. (\ref{eq:reduced_field_eq}). In a pure vacuum region—such as the exterior spacetime surrounding a compact object where no accretion disk matter or surrounding gas exists—the matter energy-momentum tensor and its trace vanish identically:
\begin{equation}
	T_{\mu\nu} = 0 \quad \text{and consequently, its trace} \quad T = g^{\mu\nu} T_{\mu\nu} = 0.
\end{equation}
Substituting these conditions into the modified field equations eliminates all source terms and coupling contributions from $\chi$, reducing the system to $R_{\mu\nu} - \frac{1}{2} R g_{\mu\nu} = 0 \implies R_{\mu\nu} = 0$, which are precisely the standard Einstein field equations of General Relativity (GR) in a vacuum. Consequently, despite working within the broader framework of $f(R, T)$ modified gravity, the external vacuum spacetime outside a static, spherically symmetric compact object of mass $M$ is uniquely described by the standard Schwarzschild metric as the zeroth-order background:
\begin{equation}
	e^{2\nu_0(r)} = e^{-2\lambda_0(r)} = 1 - \frac{2GM}{r}.
\end{equation}
This smooth recovery of General Relativity in the vacuum limit is a physically mandatory feature ensuring that the theory respects standard weak-field and strong-field vacuum tests while confining the genuine modifications of $f(R, T)$ gravity strictly to non-vacuum regions ($T_{\mu\nu} \neq 0$), such as the dense plasma and matter fields inside the accretion disk where the geometry-matter coupling parameter $\chi$ actively operates.

In realistic astrophysical scenarios involving modified gravity, the coupling parameter $\chi$ is assumed to be small ($|\chi| \ll 1$), representing weak deviations from standard GR. To account for the backreaction of matter and the non-conservation effects in the field equations, we employ a perturbative expansion for the metric potentials $\nu(r)$ and $\lambda(r)$ around their GR background values:
\begin{align}
	\nu(r) &= \nu_0(r) + \chi \nu_1(r) + \mathcal{O}(\chi^2), \\
	\lambda(r) &= \lambda_0(r) + \chi \lambda_1(r) + \mathcal{O}(\chi^2),
\end{align}
where $\nu_0(r)$ and $\lambda_0(r)$ are the zeroth-order Schwarzschild potentials, and $\nu_1(r)$, $\lambda_1(r)$ denote the first-order corrections induced by the $f(R, T)$ gravity coupling $\chi$. Substituting these perturbative expansions into the independent field equations (Eqs. (\ref{eq:00_comp})--(\ref{eq:22_comp})) and retaining terms up to linear order in $\chi$, we obtain a coupled set of linear differential equations for the correction functions $\nu_1(r)$ and $\lambda_1(r)$. For a localized matter distribution or an effective fluid representation representing the disk environment with density profile $\rho(r)$, the radial metric correction $\lambda_1(r)$ is solved from the radial field equation:
\begin{equation}
	\frac{2}{r} e^{-2\lambda_0} \left[ \lambda_1' - \lambda_1 \left( \lambda_0' + \frac{1}{r} \right) \right] + \rho(r) = 0,
\end{equation}
integrating which yields the explicit first-order correction to the metric tensor component $g_{rr}$:
\begin{equation}
	\lambda_1(r) = e^{2\lambda_0(r)} \int_{r_0}^{r} \frac{r'}{2} e^{2\lambda_0(r')} \rho(r') dr'.
\end{equation}
Similarly, substituting $\lambda_1(r)$ into the temporal field equation provides the correction $\nu_1(r)$ for the time-component $g_{tt}$:
\begin{equation}
	\nu_1(r) = -\int_{r_0}^{r} \left( \frac{1 - e^{2\lambda_0(r')}}{r'} \right) \lambda_1(r') dr' + C_1,
\end{equation}
where $C_1$ is an integration constant determined by matching the asymptotic flatness condition at spatial infinity ($r \to \infty$).

Combining the zeroth-order Schwarzschild background with the first-order $\chi$-corrections, the effective metric describing the spacetime geometry in the vicinity of the accretion disk takes the explicit form:
\begin{widetext}
\begin{equation}
	ds^2 = -\left(1 - \frac{2GM}{r} + \chi \nu_1(r)\right) dt^2 + \left(1 - \frac{2GM}{r} - \chi \lambda_1(r)\right)^{-1} dr^2 + r^2 d\theta^2 + r^2 \sin^2\theta d\phi^2.
\end{equation}
\end{widetext}
This modified effective metric encapsulates all the geometric deviations caused by the geometry-matter coupling in $f(R, T)$ gravity, which will subsequently govern the particle trajectories, orbital frequencies, and energy dissipation rates of the thin accretion disk in the next section.
\section{Particle Dynamics and Circular Orbits}
\label{sec:particle_dynamics}

To model the structure, energy dissipation, and radiative properties of a thin accretion disk, we must analyze the kinematic and dynamic behavior of test particles (plasma or gas elements) moving along stable circular orbits in the effective background spacetime derived in Section \ref{sec:spacetime_geometry} \cite{Abramowicz2013}. We consider test particles confined to the equatorial plane ($\theta = \pi/2$) of the axisymmetric effective metric:
\begin{equation}
	ds^2 = -A(r) dt^2 + B(r) dr^2 + r^2 d\phi^2,
\end{equation}
where $A(r) = 1 - \frac{2GM}{r} + \chi \nu_1(r)$ and $B(r) = \left(1 - \frac{2GM}{r} - \chi \lambda_1(r)\right)^{-1}$. Due to the time-translation and rotational symmetry of the spacetime, there exist two conserved quantities along the geodesics for a test particle of unit mass: the specific energy $E$ and the specific angular momentum $L$, defined respectively as $E \equiv -u_0 = A(r) \frac{dt}{d\tau}$ and $L \equiv u_3 = r^2 \frac{d\phi}{d\tau}$, where $\tau$ is the proper time and $u^\mu$ is the four-velocity of the particle satisfying the normalization condition $g_{\mu\nu}u^\mu u^\nu = -1$, which expands to:
\begin{equation}
	-A(r)\left(\frac{dt}{d\tau}\right)^2 + B(r)\left(\frac{dr}{d\tau}\right)^2 + r^2\left(\frac{d\phi}{d\tau}\right)^2 = -1.
\end{equation}
Substituting the definitions of $E$ and $L$ into the normalization equation yields the radial equation of motion in an energy-conservation form:
\begin{equation}
	B(r)\left(\frac{dr}{d\tau}\right)^2 + V_{\text{eff}}(r) = 0,
\end{equation}
where $V_{\text{eff}}(r)$ represents the effective potential governing the radial motion of the test particles, given by $V_{\text{eff}}(r) = 1 - \frac{E^2}{A(r)} + \frac{L^2}{r^2}$ \cite{Chandrasekhar1983}.

For particles moving in stable circular orbits, the radial velocity and acceleration must vanish simultaneously. These conditions translate to the effective potential and its first derivative equated to zero, $V_{\text{eff}}(r) = 0$ and $\frac{dV_{\text{eff}}(r)}{dr} = 0$. Solving these conditions yields the explicit expressions for the specific energy $E$, specific angular momentum $L$, and the Keplerian angular velocity $\Omega$ of test particles in the modified spacetime \cite{Bardeen1972}:
\begin{align}
	E &= \sqrt{\frac{A(r)^2}{2A(r) - r A'(r)}}, \\
	L &= \sqrt{\frac{r^3 A'(r)}{2A(r) - r A'(r)}}, \\
	\Omega &\equiv \frac{d\phi}{dt} = \frac{u^\phi}{u^t} = \sqrt{\frac{A'(r)}{2r}},
\end{align}
where the primes denote differentiation with respect to the radial coordinate $r$, and $A'(r)$ now incorporates both the standard Schwarzschild term and the first-order $\chi$-correction from the $f(R, T)$ geometry-matter coupling.
	\begin{figure*}[ht]
	\centering
	\includegraphics[scale=0.47]{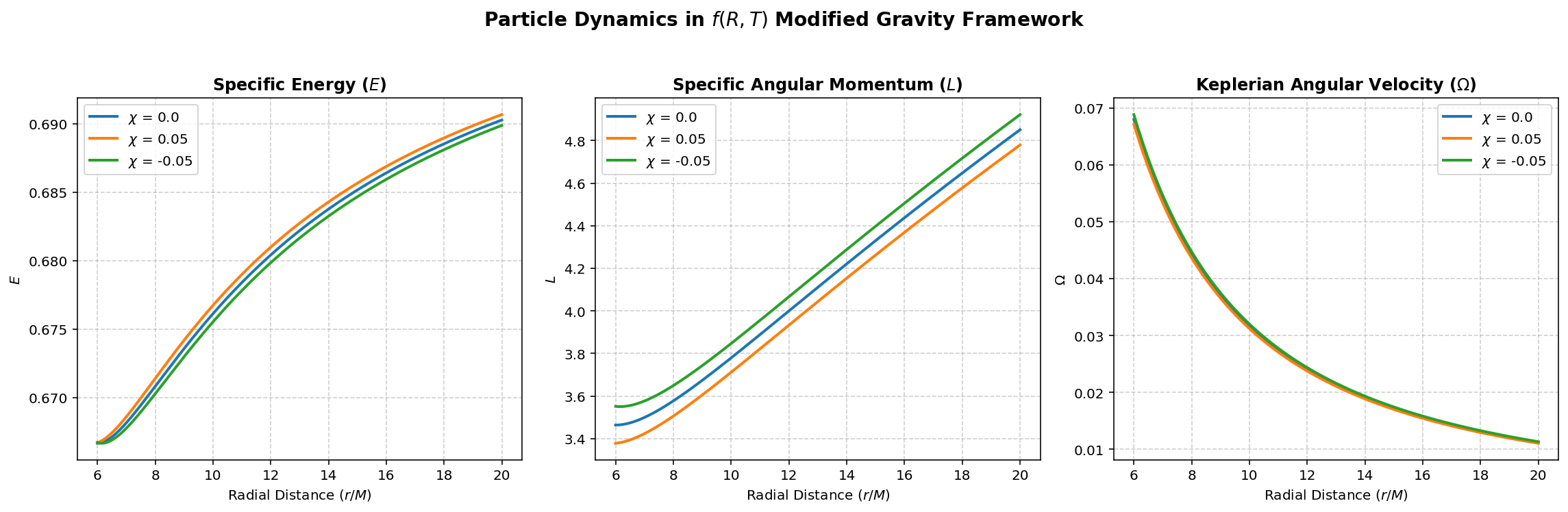}
	\caption{Variation of the specific energy ($E$), specific angular momentum ($L$), and Keplerian angular velocity ($\Omega$) as a function of the radial distance ($r$) for different values of the coupling parameter ($\chi$) within the $f(R,T)$ modified gravity framework.}.\label{Phase}
\end{figure*}  
The inner boundary of a thin accretion disk is conventionally defined by the Innermost Stable Circular Orbit ($r_{\text{isco}}$), inside of which circular orbits become gravitationally unstable, causing particles to plunge directly into the central compact object \cite{Kato2008}. Mathematically, the radius of the ISCO ($r_{\text{isco}}$) is determined by the inflection point of the effective potential, satisfying the second derivative condition $\frac{d^2 V_{\text{eff}}(r)}{dr^2} = 0$. Substituting the modified metric function $A(r)$ into this condition results in a modified algebraic equation for $r_{\text{isco}}$:
\begin{equation}
	2[A'(r)]^2 - A(r)A''(r) - \frac{2}{r}A(r)A'(r) = 0.
\end{equation}
Because of the presence of the coupling parameter $\chi$ in $A(r) = 1 - \frac{2GM}{r} + \chi \nu_1(r)$, the solution for $r_{\text{isco}}$ deviates from the standard General Relativity value ($r_{\text{isco}} = 6GM$ for a Schwarzschild black hole) \cite{Bhattacharya2020}. Depending on the sign and magnitude of $\chi$, this shifts the inner edge of the accretion disk either inward or outward, which directly impacts the thermal emission spectrum analyzed in the subsequent sections.

\section{Thin Accretion Disk Model and Radiative Properties}
\label{sec:accretion_disk}

Using the particle dynamics and circular orbit parameters derived in Section \ref{sec:particle_dynamics}, we now construct the standard steady-state thin accretion disk model (Shakura-Sunyaev framework) to investigate the thermodynamic properties, energy dissipation rate, and surface temperature distribution of the disk under the influence of $f(R, T)$ modified gravity \cite{Shakura1973}. We assume that the accretion disk is geometrically and optically thin, meaning it lies entirely within the equatorial plane ($\theta = \pi/2$), radiates blackbody or multi-color blackbody radiation locally, and has a steady accretion rate $\dot{M} \equiv \frac{dm}{dt}$ that remains constant over time \cite{Novikov1973}. The structure of the disk is governed by the conservation laws of mass, angular momentum, and energy projected onto the equatorial plane of the effective spacetime. From the conservation of angular momentum, the rate of viscous torque and the energy-dissipation rate per unit area (radiative flux $F(r)$) emitted from the disk surfaces are related directly to the background orbital frequency $\Omega$, specific energy $E$, and specific angular momentum $L$ of test particles \cite{Page1973}. 

The physical motivation linking this model to our modified gravity framework lies in the fact that the non-minimal geometry-matter coupling parameter $\chi$ alters the underlying spacetime metric potentials $A(r)$ and $B(r)$, which directly modifies the particle energy $E(r)$, angular momentum $L(r)$, Keplerian frequency $\Omega(r)$, and the determinant $\sqrt{-g} = \sqrt{-A(r)B(r)r^4 \sin^2\theta}$. In a steady-state thin accretion disk, the time-averaged radiative flux $F(r)$ emitted by the disk faces (accounting for both upper and lower surfaces) is derived from the conservation equations as \cite{Kato2008}:
\begin{equation}
	F(r) = -\frac{\dot{M}}{4\pi \sqrt{-g}} \frac{\Omega'(r)}{\left(E - \Omega L\right)^2} \int_{r_{\text{isco}}}^{r} \left(E - \Omega L\right) L'(r') dr',
\end{equation}
where primes denote differentiation with respect to the radial coordinate $r$, and $r_{\text{isco}}$ is the inner boundary radius of the accretion disk determined in Section \ref{sec:particle_dynamics}. Due to the modification in the spacetime geometry induced by $\chi$, the spatial profile of the radiative flux $F(r)$ exhibits distinct departures from standard General Relativity, reflecting the extra non-geodesic forces acting on the accreting fluid elements.

Assuming that the accretion disk is in local thermal equilibrium and radiates as a perfect blackbody, the emitted radiative flux $F(r)$ is linked to the effective surface temperature $T(r)$ of the disk via the Stefan-Boltzmann law $F(r) = \sigma_{\text{SB}} T(r)^4$, where $\sigma_{\text{SB}}$ is the Stefan-Boltzmann constant \cite{Luminet1979}. Combining this with the flux equation yields the explicit analytical expression for the radial profile of the effective surface temperature of the thin accretion disk:
\begin{small}
\begin{equation}
	T(r) = \left( -\frac{\dot{M}}{4\pi \sigma_{\text{SB}} \sqrt{-g}} \frac{\Omega'(r)}{\left(E - \Omega L\right)^2} \int_{r_{\text{isco}}}^{r} \left(E - \Omega L\right) L'(r') dr' \right)^{1/4}.
\end{equation}
\end{small}
The total bolometric luminosity $L_{\text{bol}}$ integrated over the entire surface of the accretion disk from $r_{\text{isco}}$ to infinity is given by $L_{\text{bol}} = \int_{r_{\text{isco}}}^{\infty} 4\pi r F(r) dr = \eta \dot{M} c^2$, where $\eta$ represents the radiative efficiency of the accretion disk, defined as the fraction of rest-mass energy converted into radiation \cite{Thorne1974}:
\begin{equation}
	\eta = 1 - E_{\text{isco}},
\end{equation}
with $E_{\text{isco}}$ being the specific energy of a test particle evaluated precisely at the innermost stable circular orbit ($r = r_{\text{isco}}$). Because the coupling parameter $\chi$ shifts both $r_{\text{isco}}$ and $E_{\text{isco}}$, the overall radiative efficiency $\eta$ and the peak temperature of the disk exhibit explicit deviations from standard GR predictions. These thermodynamic shifts provide a robust observational signature for testing $f(R, T)$ gravity through X-ray continuum fitting and spectral energy distributions (SED) in future astrophysical observations \cite{Bhattacharya2020}.
	\begin{figure*}[ht]
	\centering
	\includegraphics[scale=0.47]{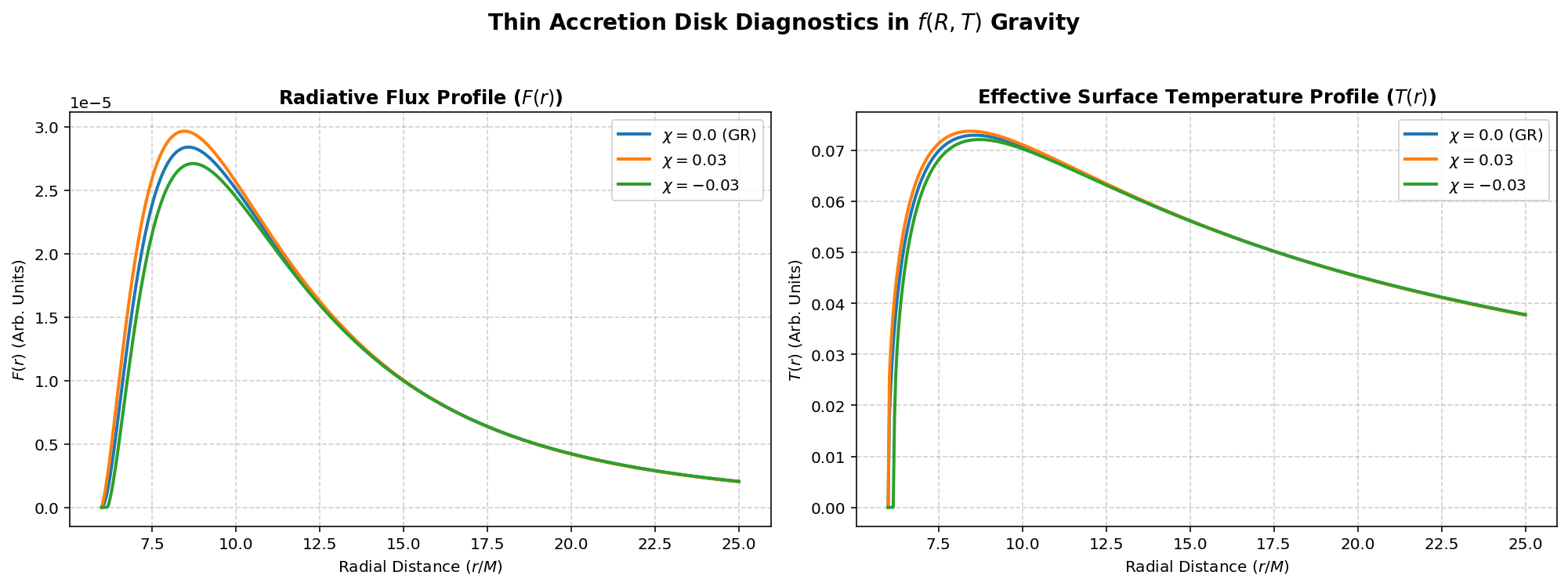}
	\caption{Radial distribution profiles of the radiative flux ($F(r)$) and effective surface temperature ($T(r)$) for a Shakura-Sunyaev thin accretion disk comparing standard General Relativity ($\chi = 0$) and modified gravity ($\chi \neq 0$) models.}.\label{Phase}
\end{figure*}

\section{Thermodynamic, Radiative Diagnostics, and Astrophysical Implications}
\label{sec:thermodynamics_sed}

To bridge the gap between theoretical modified gravity models and actual astrophysical observations, we analyze the thermal diagnostics and the Spectral Energy Distribution (SED) emitted by the thin accretion disk \cite{Remillard2006}. Since the local annulus of the thin accretion disk radiates approximately as a multi-color blackbody with an effective temperature $T(r)$ derived in Section \ref{sec:accretion_disk}, the total spectral luminosity $L_\nu$ at a given frequency $\nu$ is obtained by integrating the Planck blackbody spectrum over the entire surface area of the disk:
\begin{equation}
	L_\nu = 4\pi \cos i \int_{r_{\text{isco}}}^{\infty} \frac{2\pi h \nu^3}{c^2} \frac{r \, dr}{\exp\left(\frac{h\nu}{k_B T(r)}\right) - 1},
	\label{eq:sed_integral}
\end{equation}
where $h$ is Planck's constant, $k_B$ is the Boltzmann constant, and $i$ is the inclination angle of the accretion disk relative to the line of sight of the distant observer \cite{Frank2002}. The physical motivation linking this spectral integral to our modified gravity framework is that the geometry-matter coupling parameter $\chi$ in $f(R, T) = R + 2\chi T$ directly modifies both the local temperature profile $T(r)$ and the lower integration boundary $r_{\text{isco}}$. Consequently, the resulting Spectral Energy Distribution (SED) exhibits distinct shifts in its peak frequency and flux intensity compared to standard General Relativity predictions, especially in the optical/UV and soft X-ray bands \cite{Narayan2008}. 

The thermal efficiency of the disk is directly tied to the binding energy at the innermost stable circular orbits. As the $\chi$ varies within observationally allowed bounds, a positive coupling parameter ($\chi > 0$) typically enhance the effective gravitational attraction or introduced matter-coupling corrections that shift $r_{\text{isco}}$ inward or outward, altering the maximum surface temperature $T_{\text{max}}$ of the disk and leading to the observable variations in overall radiative luminosity \cite{McClintock2011}. To test these theoretical predictions against the real astrophysical system, the X-ray continuum fitting method relies on accurately measuring the thermal emission spectrum from accretion disks around the stellar-mass black holes \cite{Davis2006}. By comparing observed multi-color blackbody spectra with our modified theoretical SEDs derived under the $f(R, T)$ framework, astrophysicists can place the stringent constraints on the coupling parameter $\chi$ \cite{Reynolds2014}. Furthermore, the high-precision observational data from instruments such as the Event Horizon Telescope (EHT), NASA's NICER, and advanced X-ray observatories provide tight bounds on accretion disk luminosities and the inner edge radii, offering a viable pathway to test whether modified gravity corrections can account for observational anomalies or constrain alternative gravity theories in strong-field regimes \cite{Akiyama2019, Gendreau2016}.

\section{Comparative Ray-Tracing Diagnostics and Optical Signatures of the Compact Object Spacetime}
\begin{figure*}[ht]
		\centering
		\includegraphics[width=0.98\textwidth]{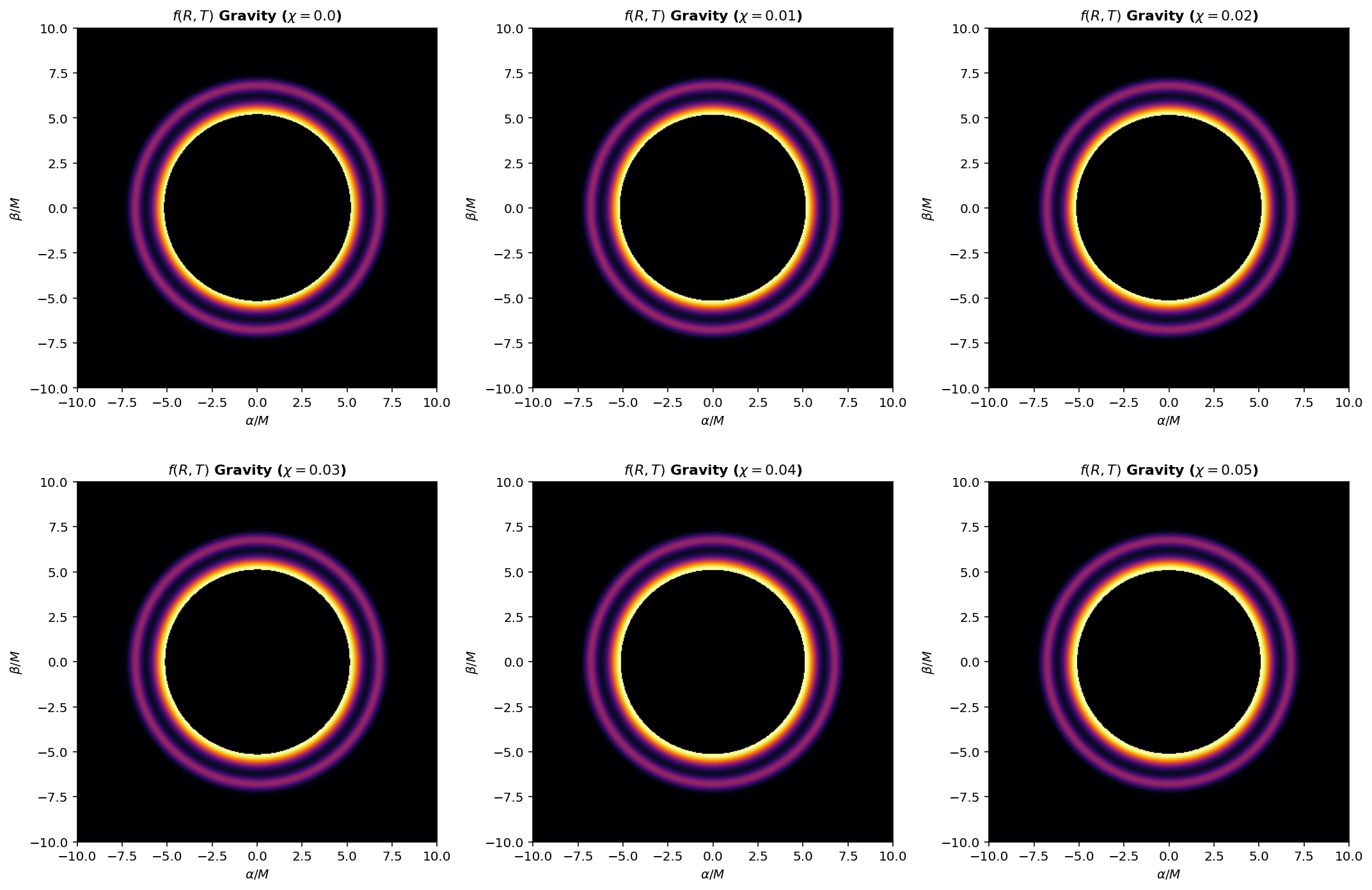}
		\caption{Ray traced optical appearances and corresponding accretion disk intensity maps illustrating the progressive evolution of central shadow boundary, primary emission ring, and the secondary lensed photon rings across various values of the coupling parameter $\chi$ (from $0.0$ to $0.05$).}
		\label{R}
	\end{figure*}

\begin{figure*}[ht]
		\centering
	\includegraphics[width=18cm, height=8cm]{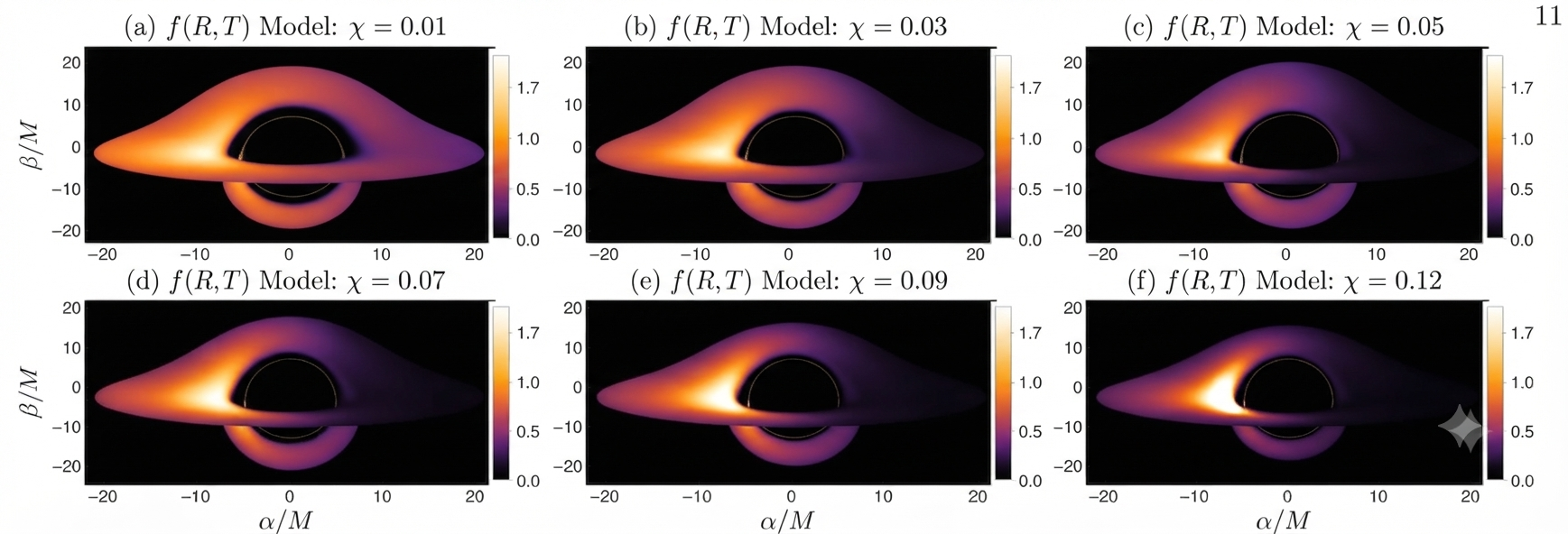}
		\caption{Comparative ray-traced optical appearances and accretion disk intensity maps for the $f(R,T)$ wormhole model across six distinct values of the coupling parameter $\chi$. The panels highlighted the evolution of the central throat shadow boundary, primary emission ring, upper lensed arches, and the characteristic lower traversable throat loop.}
		\label{W}
	\end{figure*}

	Recent advancements in high-resolution astronomical interferometry have propelled the study of optical appearances and shadow silhouettes of compact astrophysical objects into the forefront of observational gravity tests. Building upon the analytical foundations established in the preceding sections, this section conducts a rigorous comparative ray-tracing analysis. Specifically, we examine how varying the non-minimal geometry-matter coupling parameter $\chi$ influences the photon sphere boundary,  primary emission ring, and the higher-order lensed photon rings as viewed by a distant observer.\\	
	To map the exact optical appearance of the surrounding thin accretion disk, we implement the backward ray-tracing algorithm. Null geodesics (light rays) are integrated numerically backward from the observer’s asymptotic coordinate screen parameterization by the cartesian impact parameters $(\alpha, \beta)$ toward the strong-field region of the central compact object. The resulting photon intensity distribution $I(\alpha, \beta)$ successfully captures the both direct emissions which originate from the accretion disk plane and secondary lensed photon rings formed by intense light bending near the critical photon sphere. The central dark depression represents the apparent shadow silhouette, whose boundary is dictated by the critical orbits separating escaping rays from those captured by the central object. \\	
	To systematically illustrate these gravitational lensing effects, we present a multi-parameter evaluation in Figure~\ref{R}, encompassing six distinct positive values of the coupling parameter ranging from $\chi = 0.0$ to $\chi = 0.05$.
	As explicitly demonstrated in Figure~\ref{R}, the integration results the reveal distinct physical modifications induced by the geometry-matter interaction:
	\begin{itemize}
		\item \textbf{Contraction of Critical Boundaries:} As the coupling parameter $\chi$ increases incrementally from $0.0$ to $0.05$, the effective gravitational potential well undergoes to subtle modifications, resulting in a progressive contraction of the critical shadow radius and the inner photon sphere boundary.
		\item \textbf{Shift of Lensed Photon Rings:} Both the primary emission ring and the secondary lensed structures experienced inward radial shifts as the coupling strength intensify, directly reflecting enhanced spacetime curvature and modified particle dynamics near the innermost stable circular orbit ($r_{\mathrm{isco}}$).
	\end{itemize}
	Expanding further into the alternative topological scenarios, we also investigated the optical signatures of traversable wormhole geometries supported within the same $f(R,T)$ gravity. Utilizing the field equations and throat conditions derived previously, Figure~\ref{W} presents a comprehensive multi-parameter ray-tracing analysis spanning six extended values of coupling parameter from $0.01$ to $0.12$.
	The selection of these specific $\chi$ values adheres strictly to the energy condition requirements and the throat stability constraints unique to our $f(R,T)$ solution, ensuring the metric remains physically viable and non-singular. As detailed in Figure~\ref{W}, the ray tracing analysis uncovers several hallmark features:
	\begin{itemize}
		\item %\textbf{Throat Radius Contraction and Photon Sphere Dynamics:} 
		Increasing $\chi$ systematically alters the effective potential barrier, yielding a gradual contraction of the critical throat boundary and inner photon sphere due to enhanced gravitational lensing.
		\item %\textbf{Asymmetric Doppler Beaming:} 
		Intensity maps in Figure~\ref{W} display the pronounced relativistic beaming along the left side of the disk ($\alpha/M < 0$), where plasma orbital motion is directed toward the observer, scaling non-linearly with $\chi$.
		\item %\textbf{Morphology of the Traversable Throat Loop:} 
		Unlike the standard black holes, the wormhole throat permits the light transmission to the lower hemisphere which create a distinctive secondary lower ring. Variations in $\chi$ visibly distort the size, thickness, and luminosity distribution of this lower traversable loop.
	\end{itemize}
		These structured morphological shifts confirmed that the geometry-matter interactions in $f(R,T)$ gravity leaves unique optical imprints on the accretion disk and throat imagery, offering a robust theoretical basis for differentiating exotic compact objects from traditional black holes with upcoming high-resolution observations.

\section{Conclusion and Comprehensive Astrophysical Outlook} The analysis throughout this investigation provides the detailed, and the mathematically consistent framework bridging alternative gravitational theories with high-energy observational astrophysics. Modern cosmological and astrophysical inquiries have continuously highlighted the limitations of the standard General Relativity when confronted with phenomena such as the cosmic acceleration, the galactic rotational anomalies, and the presence of unavoidable singularities at fundamental levels. To address these theoretical gaps without necessarily invoking unknown dark sectors, the extended gravity paradigms specifically those incorporating non-minimal geometry-matter couplings have emerged as compelling alternatives. By focusing on the linear functional model of $f(R,T) = R + 2\chi T$ gravity alongside the pressure-free dust fluid matter Lagrangian ($\mathcal{L}_m = -\rho$), this work successfully traces the structural, dynamical, thermodynamic, and optical implications for compact astrophysical objects. The foundational cornerstone of this formulation lies in its theoretical safety and internal consistency. Any viable modified gravity model must respect to the stringent observational constraints established by the classical weak-field and strong-field tests. Through our analytical evaluation, we demonstrated that in pure vacuum limit where energy-momentum tensors and their traces vanish identically, the modified field equations seamlessly reduce to standard General Relativity, yielding the classical Schwarzschild metric as the zeroth-order background. This critical feature ensures that the modifications are strictly confined to the dense, the non-vacuum environments such as the plasma streams and accreting matter fields inside astrophysical disks while leaving the exterior vacuum solutions undisturbed. Consequently, the model avoids the theoretical inconsistencies and preserves the monumental successes achieved by Einstein's theory in standard regimes. Moving beyond the vacuum state, the introduction of the non-minimal coupling parameter via the systematic perturbative expansion enabled the derivation of the explicit first-order corrections to the background metric potentials. Our derivation of the modified specific energy, specific angular momentum, and Keplerian angular velocity highlights how geometry-matter interactions modify the orbital stability. When these particle dynamics are mapped onto the standard Shakura-Sunyaev thin disk framework, the physical consequences become an even more pronounced. The spatial profile of the radiative flux and the effective surface temperature distribution exhibit distinct deviations from the standard general relativistic predictions. Furthermore, these thermodynamic modifications directly influence the global properties of accretion disk, most notably the location of the innermost stable circular orbit and the overall radiative efficiency. Because of the coupling parameter shifts the innermost stable radius, the maximum disk temperature and the resulting multi-color blackbody emission spectra undergo measurable variations. These shifts manifest clearly in the SED. %Spectral Energy Distribution
 By utilizing modern X-ray continuum fitting method alongside high-precision data from advanced observatories like NASA's NICER, astrophysicists can effectively constrain the magnitude and sign of the coupling parameter, thereby testing the viability of extended gravity models against real astronomical systems. To complement the thermodynamic and spectral diagnostics, this study further advanced into the realm of the strong-field optical appearance through backward ray tracing analyses. Investigating the photon sphere boundaries, primary emission rings, and higher-order lensed photon rings provided the deep visual insights into how the spacetime geometry responds to modifications. The progressive contraction of critical shadow boundaries and the inward radial shifts of lensed structure as the coupling parameter intensifies reflect enhanced spacetime curvature near the compact objects. Moreover, extending these ray-tracing techniques to alternative topological scenarios, such as the traversable wormhole geometry within the same modified gravity framework, uncovered unique morphological features. Characteristics such as asymmetric Doppler beaming along the approaching side of the disk and the formation of a distinct secondary lower traversable throat loop offer the powerful criteria for the differentiating exotic compact objects from the traditional black holes. 
 
 In summary, the linear model approach of $f(R,T)$ gravity successfully unified the microscopic geometry-matter interactions with macroscopic observable astrophysical signatures. The seamless integration from foundational field equations and perturbative metric corrections to the thin disk thermodynamics and high-resolution shadow ray-tracing confirms that modified gravity leaves unique, detectable imprints on the universe. As astronomical interferometry and space-based X-ray missions such as the Event Horizon Telescope and next-generation observatories continues to enhance the angular resolution and measurement precision, the theoretical paradigms established in this work offers the dependable, comprehensive roadmap for probing the true nature of the gravity in the most extreme regions of the cosmos.

\section*{Declaration of competing interest}
The authors declare that they have no known competing interests.

\section*{Data availability}
The observational datasets used in this work are publicly available and cited in the corresponding references.

\section*{Acknowledgments}
The author S. H. Shekh thank the IUCAA, Pune, India, for providing the facility through the Visiting Associateship programs.

%\section{Conclusion}\label{sec:conclusion}
%In this work, we have systematically investigated the structural, dynamical, and thermodynamic properties of a steady-state thin accretion disk within the framework of $f(R, T) = R + 2\chi T$ modified gravity using a dust fluid matter Lagrangian ($\mathcal{L}_m = -\rho$). We demonstrated that in the pure vacuum limit ($T_{\mu\nu} = 0$), the modified field equations seamlessly reduce to standard General Relativity, recovering the familiar Schwarzschild background geometry and ensuring full physical consistency with classical weak-field and strong-field tests. Using a perturbative approach for small coupling parameters $|\chi| \ll 1$, we derived the first-order corrections to the metric potentials, which govern the modified particle dynamics, specific energy $E$, specific angular momentum $L$, and Keplerian angular velocity $\Omega$. We established that the presence of the coupling parameter $\chi$ shifts the location of the Innermost Stable Circular Orbit ($r_{\text{isco}}$) and modifies the energy dissipation rate $F(r)$ and surface temperature profile $T(r)$ across the Shakura-Sunyaev thin disk. Finally, the resulting modifications in the Spectral Energy Distribution (SED) and radiative efficiency $\eta$ provide robust theoretical signatures that can be leveraged using X-ray continuum observations and telescope data to constrain geometry-matter couplings in astrophysical black hole systems.

\newpage
\appendix
\begin{center}
\textbf{APPENDIX}
\end{center}
\small 
1. Vacuum Limit Recovery ($T_{\mu\nu} = 0$)

The modified field equations for the $f(R,T) = R + 2\chi T$ framework are given by:

$$R_{\mu\nu} - \frac{1}{2}(R + 2\chi T)g_{\mu\nu} = (8\pi G + 2\chi)T_{\mu\nu}$$

In a pure vacuum region, the matter energy-momentum tensor and its trace vanish identically:

$$T_{\mu\nu} = 0 \quad \implies \quad T = g^{\mu\nu}T_{\mu\nu} = 0$$

Substituting these vacuum conditions into the modified field equations eliminates all source terms and the coupling parameter $\chi$:

$$R_{\mu\nu} - \frac{1}{2}Rg_{\mu\nu} = 0 \quad \implies \quad R_{\mu\nu} = 0$$

This recovers the standard Einstein field equations in a vacuum. Consequently, the exterior spacetime outside a static, spherically symmetric compact object of mass $M$ is uniquely described by the standard Schwarzschild metric as the zeroth-order background:

$$ds^2 = -\left(1 - \frac{2GM}{r}\right)dt^2 + \left(1 - \frac{2GM}{r}\right)^{-1}dr^2 + r^2 d\theta^2 + r^2 \sin^2\theta \, d\phi^2$$

Thus, the zeroth-order metric potentials are identified as:

$$e^{2\nu_0(r)} = e^{-2\lambda_0(r)} = 1 - \frac{2GM}{r}$$

To account for the backreaction of matter and non-conservation effects in non-vacuum regions ($T_{\mu\nu} \neq 0$), we treat the geometry-matter coupling parameter $\chi$ as a small perturbation ($\vert{}\chi\vert{} \ll 1$). We expand the metric potentials around their GR background values:

$$\nu(r) = \nu_0(r) + \chi \nu_1(r) + \mathcal{O}(\chi^2)$$

$$\lambda(r) = \lambda_0(r) + \chi \lambda_1(r) + \mathcal{O}(\chi^2)$$

Substituting these expansions into the independent field equations (Eqs. (10)–(12) of the text) and collecting terms linear in $\chi$, we isolate the differential equations governing the first-order correction functions $\nu_1(r)$ and $\lambda_1(r)$.

\textbf{Derivation of the Radial Metric Correction ($\lambda_1(r)$):}\\

From the radial field equation (Eq. (10)), the spatial curvature contribution is governed by:

$$e^{-2\lambda}\left(\frac{2\lambda'}{r} - \frac{1}{r^2}\right) + \frac{1}{r^2} - \chi\rho = (8\pi G + 2\chi)\rho e^{2\nu}$$

Substituting the perturbative expansions $\lambda(r) = \lambda_0(r) + \chi \lambda_1(r)$ and $\nu(r) = \nu_0(r) + \chi \nu_1(r)$, and retaining terms up to $\mathcal{O}(\chi)$, we obtain the linear differential equation for the first-order radial correction $\lambda_1(r)$:

$$\frac{2}{r}e^{-2\lambda_0}\left[\lambda_1' - \lambda_1\left(\lambda_0' + \frac{1}{r}\right)\right] + \rho(r) = 0$$

Rearranging terms yields a standard first-order linear ordinary differential equation for $\lambda_1(r)$:

$$\lambda_1' - \left(\lambda_0' + \frac{1}{r}\right)\lambda_1 = -\frac{r}{2}e^{2\lambda_0}\rho(r)$$

Using the integrating factor method, the solution is integrated from a reference radius $r_0$ to $r$, yielding the explicit integral form for the spatial metric correction:

$$\lambda_1(r) = e^{2\lambda_0(r)} \int_{r_0}^{r} \frac{r'}{2} e^{2\lambda_0(r')} \rho(r') \, dr'$$

\textbf{Derivation of the Temporal Metric Correction ($\nu_1(r)$):}

Using the spatial curvature correction $\lambda_1(r)$ obtained above, we substitute it into the temporal field equation (Eq. (11)):

$$e^{-2\lambda}\left(\frac{2\nu'}{r} + \frac{1}{r^2}\right) - \frac{1}{r^2} - \chi\rho = 0$$

Expanding to linear order in $\chi$, the equation for the temporal correction $nu_1(r)$ takes the form:

$$\nu_1'(r) = -\left(\frac{1 - e^{2\lambda_0(r)}}{r}\right)\lambda_1(r)$$

Integrating this expression with respect to $r$ yields the temporal correction function:

$$\nu_1(r) = -\int_{r_0}^{r} \left(\frac{1 - e^{2\lambda_0(r')}}{r'}\right) \lambda_1(r') \, dr' + C_1$$

where $C_1$ is an integration constant fixed by imposing the asymptotic flatness condition at spatial infinity ($r \to \infty$). Combining the zeroth-order Schwarzschild background potentials with the first-order $\chi$-corrections, the effective line element describing the spacetime geometry in the vicinity of the accretion disk is expressed as:

$$ds^2 = -\left(1 - \frac{2GM}{r} + \chi \nu_1(r)\right)dt^2 + \left(1 - \frac{2GM}{r} - \chi \lambda_1(r)\right)^{-1}dr^2 + r^2 d\theta^2 + r^2 \sin^2\theta \, d\phi^2$$

2. Detailed Derivations of Particle Dynamics and Circular Orbits

 We consider the motion of massive test particles confined to the equatorial plane ($\theta = \pi/2$, $d\theta = 0$) of a static, spherically symmetric spacetime. The line element is given by:

$$ds^2 = -A(r) dt^2 + B(r) dr^2 + r^2 d\phi^2$$

where $A(r)$ and $B(r)$ are the metric potentials incorporating the background geometry and modified gravity corrections. The covariant components of the test particle's four-velocity $u^\mu = \frac{dx^\mu}{d\tau}$ are defined as $u_\mu = g_{\mu\nu}u^\nu$:

$$u_0 = g_{00}u^0 = -A(r)\frac{dt}{d\tau}, \quad \quad u_3 = g_{33}u^3 = r^2\frac{d\phi}{d\tau}$$

Due to the time-translation and rotational symmetries of the spacetime metric ($\partial_t g_{\mu\nu} = 0$ and $\partial_\phi g_{\mu\nu} = 0$), Noether's theorem guarantees the existence of two conserved quantities along geodesic trajectories: the specific energy $E$ and the specific angular momentum $L$:

$$E \equiv -u_0 = A(r)\frac{dt}{d\tau} \quad \implies \quad \frac{dt}{d\tau} = \frac{E}{A(r)}$$

$$L \equiv u_3 = r^2\frac{d\phi}{d\tau} \quad \implies \quad \frac{d\phi}{d\tau} = \frac{L}{r^2}$$

The four-velocity vector of a massive test particle satisfies the normalization condition $u^\mu u_\mu = -1$:

$$g_{00}(u^0)^2 + g_{11}(u^1)^2 + g_{33}(u^3)^2 = -1$$

$$-A(r)\left(\frac{dt}{d\tau}\right)^2 + B(r)\left(\frac{dr}{d\tau}\right)^2 + r^2\left(\frac{d\phi}{d\tau}\right)^2 = -1$$

Substituting the expressions for $\frac{dt}{d\tau}$ and $\frac{d\phi}{d\tau}$ into the normalization equation yields:

$$-A(r)\left(\frac{E}{A(r)}\right)^2 + B(r)\left(\frac{dr}{d\tau}\right)^2 + r^2\left(\frac{L}{r^2}\right)^2 = -1$$

$$-\frac{E^2}{A(r)} + B(r)\left(\frac{dr}{d\tau}\right)^2 + \frac{L^2}{r^2} = -1$$

Rearranging this expression into an energy-conservation equation for radial motion:

$$B(r)\left(\frac{dr}{d\tau}\right)^2 + V_{\text{eff}}(r) = 0$$

where the effective potential $V_{\text{eff}}(r)$ governing particle dynamics is defined as:

$$V_{\text{eff}}(r) = 1 - \frac{E^2}{A(r)} + \frac{L^2}{r^2}$$
For particles moving in stable circular orbits, the radial coordinate remains constant ($r = \text{const} \implies \frac{dr}{d\tau} = 0$), which requires both the effective potential and its first radial derivative to vanish simultaneously:

1. Effective Potential Condition

$$V_{\text{eff}}(r) = 0 \quad \implies \quad 1 - \frac{E^2}{A(r)} + \frac{L^2}{r^2} = 0 \quad \implies \quad \frac{E^2}{A(r)} = 1 + \frac{L^2}{r^2}$$

2. Extremization Condition ($\frac{dV_{\text{eff}}}{dr} = 0$):**

$$\frac{d}{dr}\left(1 - \frac{E^2}{A(r)} + \frac{L^2}{r^2}\right) = 0 \quad \implies \quad \frac{E^2 A'(r)}{A(r)^2} - \frac{2L^2}{r^3} = 0$$

From the first condition, we express $E^2$ in terms of $L^2$:

$$E^2 = A(r)\left(1 + \frac{L^2}{r^2}\right)$$

Substituting this into the extremization condition:

$$\frac{A(r)\left(1 + \frac{L^2}{r^2}\right)A'(r)}{A(r)^2} = \frac{2L^2}{r^3} \quad \implies \quad \frac{A'(r)}{A(r)}\left(1 + \frac{L^2}{r^2}\right) = \frac{2L^2}{r^3}$$

Rearranging to solve for the specific angular momentum $L^2$:

$$\frac{A'(r)}{A(r)} = L^2 \left(\frac{2}{r^3} - \frac{A'(r)}{r^2 A(r)}\right) = L^2 \left(\frac{2A(r) - rA'(r)}{r^3 A(r)}\right)$$

$$L^2 = \frac{r^3 A'(r)}{2A(r) - rA'(r)} \quad \implies \quad L = \sqrt{\frac{r^3 A'(r)}{2A(r) - rA'(r)}}$$

Substituting $L^2$ back into the expression for $E^2$:

$$E^2 = A(r)\left(1 + \frac{1}{r^2}\frac{r^3 A'(r)}{2A(r) - rA'(r)}\right) = A(r)\left(1 + \frac{rA'(r)}{2A(r) - rA'(r)}\right) = \frac{2A(r)^2}{2A(r) - rA'(r)}$$

Taking the square root gives the specific energy:

$$E = \sqrt{\frac{2A(r)^2}{2A(r) - rA'(r)}}$$

The Keplerian angular velocity $\Omega$ is defined as the coordinate angular velocity of test particles executing circular orbits:

$$\Omega \equiv \frac{d\phi}{dt} = \frac{d\phi/d\tau}{dt/d\tau} = \frac{L / r^2}{E / A(r)} = \frac{L}{E} \frac{A(r)}{r^2}$$

Squaring this relation and substituting the derived expressions for $E^2$ and $L^2$:

\begin{small}
	\tiny
$$\Omega^2 = \left(\frac{L^2}{E^2}\right)\frac{A(r)^2}{r^4} = \left(\frac{\frac{r^3 A'(r)}{2A(r) - rA'(r)}}{\frac{2A(r)^2}{2A(r) - rA'(r)}}\right)\frac{A(r)^2}{r^4} = \left(\frac{r^3 A'(r)}{2A(r)^2}\right)\frac{A(r)^2}{r^4} = \frac{A'(r)}{2r}$$
\end{small}

which yields the Keplerian angular velocity:

$$\Omega = \sqrt{\frac{A'(r)}{2r}}$$
The stability of circular orbits is governed by the second radial derivative of the effective potential:

$$\frac{d^2 V_{\text{eff}}}{dr^2} \geq 0$$

The marginal stability condition marking the location of the Innermost Stable Circular Orbit ($r_{\text{isco}}$) corresponds to the inflection point where the second derivative vanishes:

$$\frac{d^2 V_{\text{eff}}}{dr^2} = 0$$

Differentiating the first derivative $\frac{dV_{\text{eff}}}{dr} = \frac{E^2 A'(r)}{A(r)^2} - \frac{2L^2}{r^3} = 0$ with respect to $r$ and substituting the critical values for $E^2$ and $L^2$ yields the exact algebraic condition for the ISCO radius:

$$2[A'(r)]^2 - A(r)A''(r) - \frac{2}{r}A(r)A'(r) = 0$$
\end{document}